\documentclass[prb,aps,twocolumn,showpacs,floatfix]{revtex4}

\usepackage{graphicx}

\begin{document}

  \title{On the absence of a spiral magnetic order in Li$_2$CuO$_2$ \\
with  one-dimensional CuO$_2$ ribbon chains}

  \author{H. J. Xiang}
  \affiliation{Department of Chemistry, North Carolina State University, Raleigh,
    North Carolina 27695-8204}

  \author{C. Lee}
  \affiliation{Department of Chemistry, North Carolina State University, Raleigh,
    North Carolina 27695-8204}

  \author{M.-H. Whangbo}
  \thanks{Corresponding author. E-mail: mike\_whangbo@ncsu.edu}

  \affiliation{Department of Chemistry, North Carolina State University, Raleigh,
    North Carolina 27695-8204}

\date{\today}

\begin{abstract}
On the basis of first principles density functional theory electronic
structure calculations as well as classical spin analysis,
we explored why the magnetic oxide Li$_2$CuO$_2$, consisting of CuO$_2$ ribbon
chains made up of edge-sharing CuO$_4$ squares, does not exhibit a
spiral-magnetic order. Our work shows that, due to the
next-nearest-neighbor interchain interactions, the observed collinear
magnetic structure becomes only
slightly less stable than the spin-spiral ground state, 
and many states become nearly degenerate in energy with the observed
collinear structure. This suggests that the collinear magnetic
structure of Li$_2$CuO$_2$ is a consequence of order-by-disorder induced by
next-nearest-neighbor interchain interactions.
\end{abstract}


\maketitle
Copper oxides with CuO$_2$ ribbon chains made up of edge-sharing CuO$_4$ squares have
one-dimensional chains of spin-$\frac{1}{2}$ Cu$^{2+}$ ions, and
exhibit unique physical properties.  
LiCu$_2$O$_2$ \cite {Park2007} and LiCuVO$_4$ \cite {Naito2007} show
ferroelectricity when their CuO$_2$ ribbon chains undergo a
spiral-magnetic order at low temperatures.
For a chain of spin-$\frac{1}{2}$ ions, a spin spiral structure is predicted
when the nearest-neighbor (NN) ferromagnetic (FM) spin exchange $J_1$ and
the next-nearest-neighbor (NNN) antiferromagnetic (AFM) spin exchange
$J_2$ satisfy the condition $|J_2/J_1| > 0.25$, while an FM structure
is predicted if $|J_2/J_1| < 0.25$.\cite{Bursill1995}
The copper oxide Li$_2$CuO$_2$ also consists of CuO$_2$ ribbon chains, but has
a different magnetic structure.
A neutron powder diffraction study of Li$_2$CuO$_2$ at 1.5 K showed a
collinear magnetic structure in which the spins of each CuO$_2$ chain has
an FM arrangement with Cu moments perpendicular to the plane of the
CuO$_2$ ribbon and the arrangement between adjacent FM chains is AFM
\cite{Sapina1990} (hereafter this magnetic structure will be referred
to as the AFM-I state). Thus, to explain this collinear magnetic
structure, one might expect $|J_2/J_1| < 0.25$ for the CuO$_2$ chains
of Li$_2$CuO$_2$. Indeed, Graaf {\it et al.} obtained $|J_2/J_1| = 0.15$
on the basis of first principles electronic structure calculations
using the embedded cluster model.\cite{Graaf2002}
However, the CuO$_2$ ribbon chains of Li$_2$CuO$_2$ are similar in structure to
those of LiCu$_2$O$_2$ and LiCuVO$_4$, so that $|J_2/J_1| > 0.25$
would have been expected. If $|J_2/J_1| > 0.25$, one needs to ask why a
spiral magnetic order does not occur in Li$_2$CuO$_2$. In addition, more
than two spin exchange interactions are necessary to describe the
magnetic structure of Li$_2$CuO$_2$, and the nature and magnitude of these
interactions are not unequivocal.\cite{Boehm1998, Mizuno1999} Another
puzzle concerning Li$_2$CuO$_2$ is that it undergoes a phase transition
below  $\sim$2.4 K to a state believed to be 
a spin canted state.\cite{Staub2000, Ortega1998, Chung2003} So far, the origin and the
nature of this phase transition remain unclear.

The spiral magnetic order of LiCu$_2$O$_2$ and LiCuVO$_4$ is a consequence of
the spin frustration associated with the NN FM and NNN AFM
interactions in their CuO$_2$ chains. A collinear magnetic order can
occur as a consequence of order-by-disorder,\cite{order, Villain}
which occurs typically in highly spin frustrated
systems.\cite{Greedan} Provided that a spin spiral state is the ground
state for 
the CuO$_2$ chains of Li$_2$CuO$_2$, one might speculate if the AFM-I state of
Li$_2$CuO$_2$ is close in energy to the spin spiral state and if Li$_2$CuO$_2$ has
a large number of nearly degenerate states around the AFM-I state. In
the present work we explore these possibilities by studying the
magnetic structure of Li$_2$CuO$_2$ on the basis of first principles density
functional theory (DFT) electronic structure calculations and carrying
out classical spin analysis with the spin exchange parameters deduced
from the DFT calculations. 

\begin{table}[htbp]
\caption{Relative energies (in meV/Cu) of the various magnetic states with
respect to the FM state obtained from GGA+U calculations with
different $U_{eff}$ values.} 
\begin{tabular}{ccccccc}
    \hline
    \hline
    $U_{eff}$ (eV)& 0  & 2  & 4  & 6 &8  & 10  \\
    \hline
    E(AF1) &-9.59 &-5.80 &-3.84 & -2.38 & -1.46 & -0.96 \\
  E(AF2) &0.69&4.89&5.76&5.82&4.70&3.67\\
  E(AF3) &-15.34&-8.01&-4.29&-1.960&-0.71&-0.18\\
  E(AF4) &-6.66&-1.00&1.23&2.23&2.22&1.88\\
  E(AF5)&1.25&5.00&5.77&5.80&4.65& 3.65\\
  \hline
    \hline
\end{tabular}
\label{table1}
\end{table}

Our DFT electronic structure calculations employed the full-potential
augmented plane wave plus local orbital method as implemented in the
WIEN2k code.\cite{wien2k} For the exchange-correlation energy
functional, the generalized gradient approximation (GGA)
\cite{Perdew1996} was employed \cite{footnote} with
$R_{MT}^{min}K_{max}$ = 7.0. To properly describe the strong electron
correlation in the 3d transition-metal oxide, the GGA plus on-site
repulsion $U$ method (GGA$+$U) was employed.\cite{Anisimov1993}
We also examined the energy of Li$_2$CuO$_2$ as a function of the magnetic
order parameter $\mathbf{q}$ by employing the non-collinear magnetism code,
WIENncm.\cite{Laskowski2004}

\begin{figure}[htbp]
  \includegraphics[width=7.5cm]{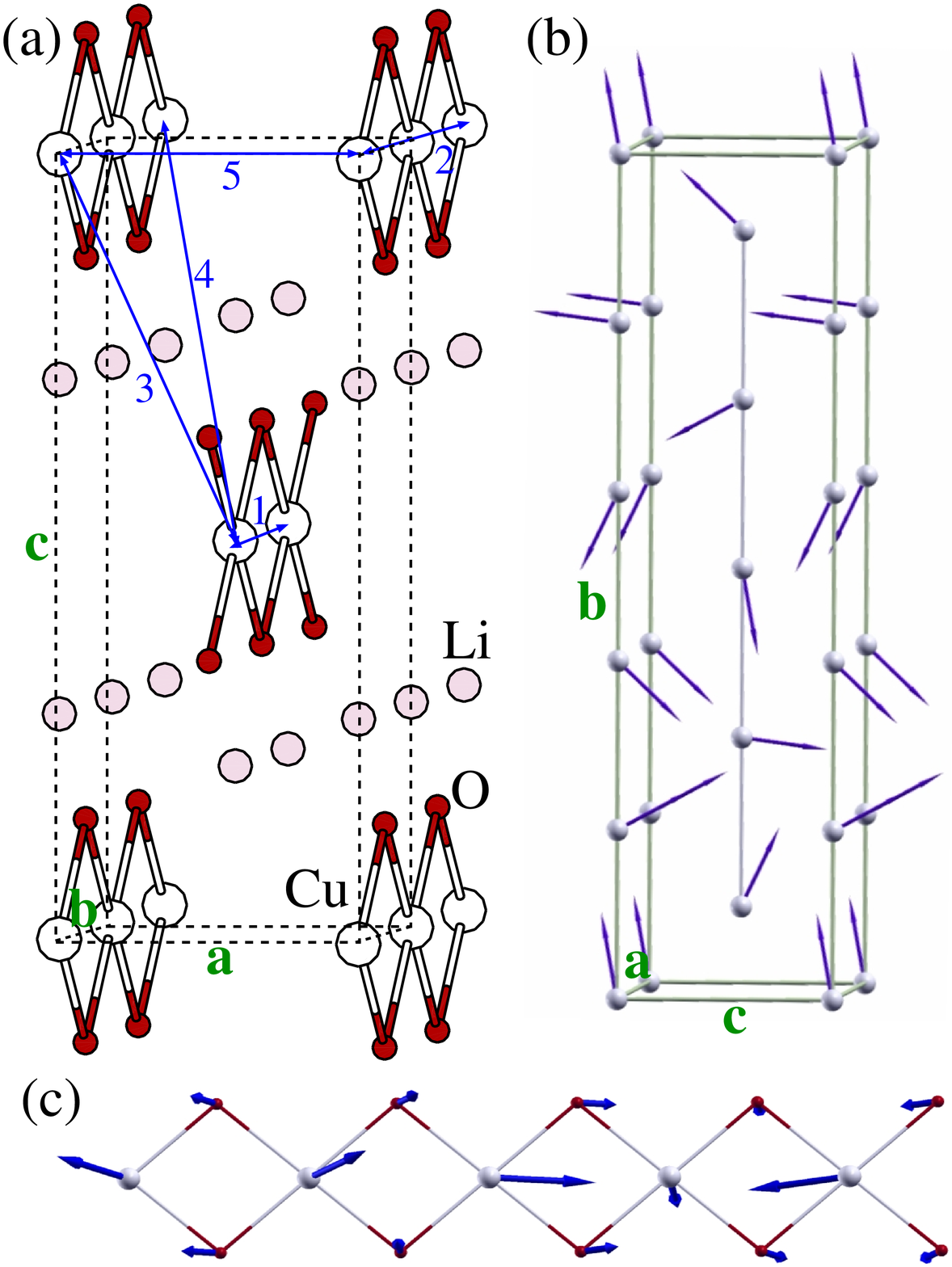}
  \caption{(Color online) (a) Crystal structure and five spin exchange paths
    $J_1- J_5$ of Li$_2$CuO$_2$. (b) Cu moments of the spin spiral ground
    state at $\mathbf{q}=(0,0.20,0)$ obtained from the GGA$+$U
    non-collinear calculation with $U_{eff} = 6$ eV. (c) Detailed view of
    the Cu and O moments of a CuO$_2$ ribbon chain in the spin spiral
    ground state shown in (b). For the purpose of illustration, the O
    moments were increased by three times.}
  \label{fig1}
\end{figure}

\begin{figure}[htbp]
  \includegraphics[width=8.0cm]{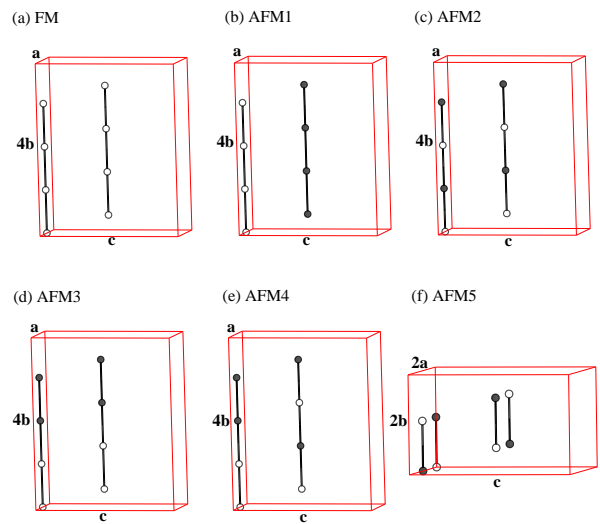}
\caption{(Color online) Schematic representations of the six ordered spin arrangements of
Li$_2$CuO$_2$ employed for GGA+U calculations to extract the five spin
exchange parameters $J_1-J_5$. The filled and empty circles the
up-spin and down-spin Cu sites, respectively.}
\label{fig2}
\end{figure}

\begin{figure}[htbp]
  \includegraphics[width=8.0cm]{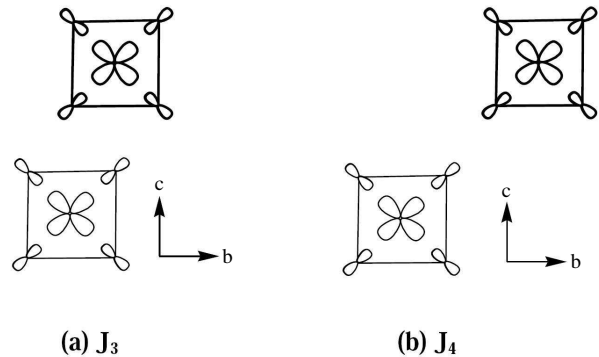}
\caption{Arrangements of the CuO$_4$ squares and their magnetic orbitals
associated with (a) the NN interchain interaction $J_3$ and (b) the NNN
interchain interaction $J_4$. The two adjacent CuO$_2$ ribbon chains differ
in their a-axis heights by a/2. The CuO$_4$ squares with different a-axis
heights are indicated by thick and thin lines.} 
\label{fig3}
\end{figure}

Li$_2$CuO$_2$ has a body centered orthorhombic structure (space group Immm
with $a$ = 3.654 \AA, $b$ = 2.860 \AA, and $c$ = 9.377 \AA),\cite{Sapina1990} where the CuO$_2$ ribbon chains run along the
b-direction (Fig.~\ref{fig1}). As depicted in Fig.~\ref{fig1}a, there are five
possible spin exchange interactions to consider; $J_1$ and $J_2$ are NN and
NNN intrachain interactions, respectively, $J_3$ and $J_4$ are NN and NNN
interchain interactions along the $c$-direction, respectively, while $J_5$
is the interchain interaction along the $a$-direction.  To evaluate the
interactions $J_1-J_5$, we calculate the relative energies of the six
ordered collinear spin states shown in Fig.~\ref{fig2} in terms of GGA+U
calculations. To see the dependence of these spin exchange
interactions on the effective on-site repulsion $U_{eff}=U-J$, our
GGA+U calculations were carried out with $U_{eff}$ ranging from 0 to
10 eV. (For 3d transition metals, U is generally less than 10 eV and
the J value is usually 1 eV.) The relative energies of the six ordered
spin states of Fig.~\ref{fig2} obtained from our GGA+U calculations are
summarized in Table~\ref{table1}. In terms of the exchange parameters $J_1-J_5$,
the energies of the six magnetic states per Cu are written as
\begin{equation}
  \begin{array}{ccl}
    E(FM) &=& (J_1 + J_2 + 4J_3 + 4J_4 + J_5)/4 \\
    E(AF1) &=& (J_1 + J_2 -4J_3 -4J_4 + J_5)/4 \\
    E(AF2) &=& (-J_1 + J_2 + J_5)/4 \\
    E(AF3) &=& (-J_2 + 2J_3 - 2J_4 + J_5)/4 \\
    E(AF4) &=& (-J_1 + J_5)/8 \\
    E(AF5) &=& (-J_1 + J_2 -J_5)/8 
  \end{array}
\end{equation}
Thus, by equating the energy differences of these states in terms of
the spin exchange parameters with the corresponding energy differences
in terms of the GGA+U calculations,  we obtain the values
of $J_1-J_5$ summarized in Table~\ref{table2}, where we employed the convention in
which positive and negative numbers represent AFM and FM interactions,
respectively. 
$J_5$ is very weak in agreement with Mizuno {\it et al.}.\cite{Mizuno1999} 
The NNN interchain interaction $J_4$ is much
stronger 
than the NN interchain interaction $J_3$, and this finding does not
support the assumption by Mizuno {\it et al.} that $J_3$ and $J_4$ are
similar.\cite{Mizuno1999} $J_4$ is stronger than $J_3$ because the overlap
between the magnetic orbitals, which depends on the overlap between
the O 2p orbitals of the magnetic orbitals,\cite {mapping} is much
more favorable for the path $J_4$ than for the path $J_3$
(Fig.~\ref{fig3}).  The NN intrachain interaction $J_1$ is FM
while the NNN intrachain interaction $J_2$ is AFM. 
These intrachain interactions are the same in
nature to those reported by Graaf {\it et al.}.
\cite{Graaf2002} However, our study shows that $|J_2/J_1| > 0.25$,
for all $U_{eff}$ values employed, and hence Li$_2$CuO$_2$ should have a
spin-spiral ground state as far as isolated CuO$_2$ ribbon chains are
concerned.



\begin{table}[htbp]
  \caption{Calculated exchange parameters (in meV) deduced from GGA$+$U
    calculations.} 
  \begin{tabular}{ccccccc}
    \hline
    \hline
    $U_{eff}$ (eV)& 0  & 2  & 4  & 6 &8  & 10  \\
    \hline
    $J_1$  & -10.98 & -15.58 & -15.36 & -14.02 & -10.86  & -8.31 \\
    $J_2$  & 23.91 & 15.78 & 10.44 & 7.35 & 4.48 & 3.00 \\
    $J_3$  & 1.07& 0.34& -0.04 & -0.01& -0.09 & -0.03 \\
    $J_4$  & 3.73 & 2.56 & 1.95 & 1.20 & 0.82 & 0.51 \\
    $J_5$  & -1.12 & -0.22 & -0.01 & 0.05 & 0.09 &0.05 \\
    \hline
    \hline
  \end{tabular}
  \label{table2}
\end{table}

\begin{figure}[htbp]
  \includegraphics[width=7.5cm]{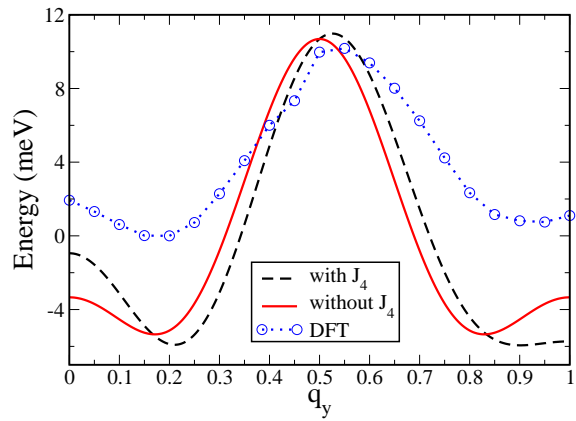}
  \caption{(Color online) E(0, $q_y$, 0) vs. (0, $q_y$, 0) relations calculated for
    Li$_2$CuO$_2$. The solid and
    dashed lines are based on the classical spin analysis 
    (solid line: only with the intrachain interactions $J_1$ and $J_2$,
    dashed line: with the intrachain interactions $J_1$ and $J_2$ as well as
    the interchain interaction $J_4$). The dotted line is based on
    non-collinear GGA$+$U calculations with $U_{eff} = 6$ eV, where the
    circles represent the calculated points.}
  \label{fig4}
\end{figure}

To see how the above prediction is affected by the NNN interchain
interaction $J_4$, we carried out a classical spin analysis based on the
Freiser method \cite{Freiser1961, Dai2004} using the three dominant
exchange parameters $J_1$, $J_2$, and $J_4$.
The spin interaction energy of an ordered spin state
with $\mathbf{q}=(2\pi q_x/a,2\pi q_y/b,2\pi q_z/c)$ can be written as 
\begin{equation}
\begin{array}{lll}
E(\mathbf{q})&=&J_4 \{ \mathrm{cos}[2\pi (q_x/2+3q_y/2+q_z/2)] \\
  && + \mathrm{cos}[2\pi (-q_x/2+3q_y/2+q_z/2)] \\
  &&+ \mathrm{cos}[2\pi (q_x/2-3q_y/2+q_z/2)] \\
  && + \mathrm{cos}[2\pi(q_x/2+3q_y/2-q_z/2)] \} \\
&&+\mathrm{cos}(2\pi q_y) J_1 + \mathrm{cos}(4\pi q_y) J_2.
\end{array}
\end{equation}
For simplicity of our discussion, we will represent $\mathbf{q}$ by ($q_x$, $q_y$, $q_z$).
This E($\mathbf{q}$) vs. $\mathbf{q}$ relation has minima along the ($0,q_y,0$)
direction. The E(0, $q_y$, 0) vs. (0, $q_y$, 0)  curves calculated with the spin exchange
parameters derived from the GGA$+$U calculations for $U_{eff} = 6$ eV are
presented in Fig.~\ref{fig4}.  The solid curve, obtained only with
the intrachain interactions $J_1$ and $J_2$, shows two minima (at $q_y=0.18$
and $q_y=0.82$) of equal energy.
The FM
state ($q_y=0.00$) and the AFM-I state ($q_y=1.00$) are identical in
energy, and are less stable than the two spin-spiral states
($q_y=0.18$ and $q_y=0.82$). These are the expected results in the
absence of the interchain interaction because $|J_2/J_1| > 0.25$. The
dashed curve, obtained with the intrachain interactions $J_1$ and $J_2$ as
well as the interchain interaction $J_4$, also shows two minima at
$q_y=0.21$ and $q_y=0.90$.
Note that the interchain interaction $J_4$ raises
the energy of the FM state while lowering that of the AFM-I state. As
a result, the E(0, $q_y$, 0) vs. (0, $q_y$, 0)   curve around $q_y=0.21$ becomes sharper while
that around $q_y=0.90 - 1.00$ is nearly flat. Both spin-spiral states
are only slightly more stable than the collinear AFM-I state. Our
calculations using the spin exchange parameters obtained with $U_{eff} <
6 $
eV show that the energy around $q_y=0.21$ becomes lower than that
around $q_y=0.90$, and both states have lower energies than the
collinear AFM-I state ($q_y=1.00$). In terms of the parameters obtained
for $U_{eff} > 6$ eV, however, the collinear AFM-I state becomes the ground
state.

Now we evaluate E(0, $q_y$, 0) vs. (0, $q_y$, 0) relations on the basis of
non-collinear GGA+U electronic structure calculations using the
WIENncm code.\cite {Laskowski2004}
In this method, the
incommensurate spiral magnetic order is simulated without resorting to
the supercell technique by using the generalized Bloch
theorem.\cite{Sandratskii1998}
 The E(0, $q_y$, 0) vs. (0, $q_y$, 0)  relation calculated for the
representative $U_{eff}$ (i.e., 6 eV), presented in Fig.~\ref{fig4} as a dotted
line, is quite similar to that found from the classical spin analysis.
An important
difference is that the non-collinear GGA$+$U calculations predict the
spin-spiral state at $q_y=0.20$ to be slightly more stable that that
at $q_y=0.95$. The spin arrangement of the spin-spiral state at
$q_y=0.20$ is illustrated in Fig.~\ref{fig1}b and ~\ref{fig1}c. In this state of zero
total spin moment, the non-collinearity of the spin arrangement occurs
not only between Cu spins but also between the O and Cu spins. Our
calculations show substantial moments on the O sites, as found in the
previous studies.\cite{Weht1998, Chung2003, Staub2000} From our
calculation with $U_{eff}=6.0$ eV, the oxygen spin moment is 0.11 $\mu_B$,
which agrees with the LDA$+$U result \cite{Mertz2005} and the
experimental value (between 0.10 and 0.12 $\mu_B$).\cite{Chung2003}
Our non-collinear GGA$+$U electronic structure
calculations with $U_{eff} > 6$ eV or with $U_{eff} < 6$ eV still show that the
ground state is a spin-spiral state. Thus, with any reasonable $U$
value, we predict a spin-spiral ground state for Li$_2$CuO$_2$.

From our non-collinear GGA$+$U calculations, the energy difference
between the spin-spiral state at $\mathbf{q}=(0,0.20,0)$ and AFM-I
state  at $\mathbf{q}=(0,1.00,0)$ is very small (Fig.~\ref{fig4}). In
the case of $U_{eff}=6$ eV, the difference 
is 1 meV/Cu and deceases with increasing $U_{eff}$. From the classical spin
analysis shown in Fig.~\ref{fig4}, this energy difference is even
smaller. As already pointed out, the E(0, $q_y$, 0) vs. (0, $q_y$, 0) curve is sharp around
$q_y=0.20$ but nearly flat around $q_y=0.90- 1.00$. As a consequence,
the states in the region of $q_y=0.90 - 1.00$ are nearly degenerate,
 and are only slightly less stable than the spin-spiral ground state
 at $q_y \sim 0.20$, namely, the density of states is much higher in
 the region of the AFM-I state than around the spin-spiral ground
 state. The latter provides a natural explanation for why the CuO$_2$
 ribbon chains of Li$_2$CuO$_2$ do not exhibit a spiral-magnetic order
 despite that the CuO$_2$ chains are very similar in structure to those
 found in LiCu$_2$O$_2$ and LiCuVO$_4$, and Li$_2$CuO$_2$ has a spin-spiral ground
 state. In short, the AFM-I structure ($q_y=1.00$) is
a collinear order arising from the occupation of many nearly
degenerate states around $q_y=0.90 - 1.10 $, and hence is an example of
order-by-disorder.\cite{order, Villain} The phase transition below 2.4 K,
  believed to be a transition to a spin canted state, might arise from
  an increased population of the spin-spiral state at
  $\mathbf{q}=(0,0.20,0)$. What distinguishes Li$_2$CuO$_2$ from LiCu$_2$O$_2$ and
  LiCuVO$_4$ is the NNN interchain interaction $J_4$, which lowers the
  energy of the states around the AFM-I state and makes them nearly
  degenerate.

Our work was supported by the Office of Basic Energy Sciences,
Division of Materials Sciences, U. S. Department of Energy, under
Grant No. DE-FG02-86ER45259. We thank Dr. D. Dai for useful discussions.



\begin{thebibliography}{99}
\bibitem{footnote}Our LDA calculations give qualitatively similar results.
\end{thebibliography}
\end{document}